\documentclass[conference,a4paper]{IEEEtran}
\usepackage[left=1.43cm,right=1.43cm,top=1.8cm,bottom=4.21cm]{geometry}

\usepackage[algo2e]{algorithm2e} 
\usepackage{amsmath,amssymb,amsmath,amsfonts}
\usepackage{bm}
\usepackage{bbm}
\usepackage{graphicx}
\usepackage{cite}
\usepackage{epstopdf}
\usepackage{gensymb}
\usepackage{color}
\usepackage{lipsum}
\usepackage{float}
\usepackage{epstopdf}
\usepackage[mathcal]{eucal}
\usepackage{subfiles}
\usepackage[T1]{fontenc}
\usepackage{siunitx}
\usepackage{tabulary}
\usepackage{epsfig,graphics,graphicx,subfig,amssymb,amstext,amsmath,algorithm,algorithmic,wrapfig,multirow}
\usepackage{array}
\usepackage{adjustbox}
\usepackage[dvipsnames]{xcolor}
\usepackage{cite}
\usepackage{times}
\usepackage{placeins}
\usepackage{textcomp}
\usepackage[font=small]{caption}
\usepackage{flushend}
\usepackage{color, colortbl}
\usepackage{tabularx}
\usepackage{bm}
\usepackage{enumerate}
\usepackage{tikz}
\usepackage{pgfplots}
\usepackage{epstopdf}
\usetikzlibrary{shapes,arrows}
\usepackage[utf8]{inputenc}
\usepackage{mathtools}
\usepackage[utf8]{inputenc}
\usepackage{xcolor}
\usepackage{tikz}
\usetikzlibrary{chains,arrows,calc,positioning}
\usepackage{amssymb}
\usepackage{pifont}
\usepackage{soul}
\usepackage{breqn} 
\usepackage{booktabs} 
\usepackage{multirow}
\usepackage{enumitem}
\usepackage{tabulary}
\usepackage[normalem]{ulem}
\usepackage{dblfloatfix}
\usepackage{makecell}
\usepackage{lipsum}
\usepackage{amsthm}
\usepackage{comment}
\usepackage{filecontents}

\newtheoremstyle{mystyle}
  {}
  {}
  {\itshape}
  {}
  {\bfseries}
  {.}
  { }
  {}
\theoremstyle{mystyle}

\newlength \figwidth
\definecolor{bittersweet}{rgb}{1.0, 0.44, 0.37}
\definecolor{glaucous}{rgb}{0.38, 0.51, 0.71}
\definecolor{gainsboro}{rgb}{0.86, 0.86, 0.86}
\definecolor{babyblueeyes}{rgb}{0.63, 0.79, 0.95}
\definecolor{silver}{rgb}{0.75, 0.75, 0.75}
\definecolor{neoncarrot}{rgb}{1.0, 0.64, 0.26}
\definecolor{Gray}{gray}{0.9}
\definecolor{LightCyan}{rgb}{0.88,1,1}
\definecolor{BackgroundLightBlue}{rgb}{0.97,0.97,1}
\definecolor{BackgroundGray}{gray}{0.98}

\newcommand{\green}[1]{{\textcolor[rgb]{0,0.5,0}{#1}}}

\newcommand{\gio}[1]{\noindent \green{ {{$\blacktriangleright$ 
   {\textsf{[Gio]: #1}} $\blacktriangleleft$}}}}

\makeatletter

 \let\oldforeign@language\foreign@language
 \DeclareRobustCommand{\foreign@language}[1]{%
   \lowercase{\oldforeign@language{#1}}}

\def\nb0{{\mathbf{0}}}
\def\nb1{{\mathbf{1}}}

\def\ncalB{{\mathcal{B}}}

\def\ncalG{{\mathcal{G}}}

\def\ncalU{{\mathcal{U}}}

\def\sinr{\mathtt{SINR}}			

\def\calB{\mathcal{B}}

\makeatother

\IEEEoverridecommandlockouts
\begin{document}

\bstctlcite{IEEEexample:BSTcontrol}

\title{Designing Cellular Networks for UAV Corridors via Bayesian Optimization}

\author{\IEEEauthorblockN{Mohamed Benzaghta$^{\star}$, Giovanni Geraci$^{\star}$, David L\'{o}pez-P\'{e}rez$^{\sharp}$, and Alvaro Valcarce$^{\flat}$\vspace{0.1cm}
}
\\ \vspace{-0.3cm}
\normalsize\IEEEauthorblockA{$^{\star}$\emph{Univ. Pompeu Fabra, Barcelona, Spain} \enspace \enspace  $^{\sharp}$\emph{Univ. Politècnica de València, Spain} \enspace \enspace  $^{\flat}$\emph{Nokia Bell Labs, Massy, France} }
\thanks{This work was supported by the Spanish Research Agency through grant PID2021-123999OB-I00, CEX2021-001195-M, and the ``Ram\'{o}n y Cajal'' program, by the UPF-Fractus Chair, and by Generalitat Valenciana, Spain, through grants CIDEGENT PlaGenT, CIDEXG/2022/17, and Project iTENTE.} 
}

\maketitle

\begin{abstract}

As traditional cellular base stations (BSs) are optimized for 2D ground service, providing 3D connectivity to uncrewed aerial vehicles (UAVs) requires re-engineering of the existing infrastructure. In this paper, we propose a new methodology for designing cellular networks that cater for both ground users and UAV corridors based on Bayesian optimization. We present a case study in which we maximize the signal-to-interference-plus-noise ratio (SINR) for both populations of users by optimizing the electrical antenna tilts and the transmit power employed at each BS. Our proposed optimized network significantly boosts the UAV performance, with a 23.4~dB gain in mean SINR compared to an all-downtilt, full-power baseline. At the same time, this optimal tradeoff nearly preserves the performance on the ground, even attaining a gain of 1.3~dB in mean SINR with respect to said baseline. Thanks to its ability to optimize black-box stochastic functions, the proposed framework is amenable to maximize any desired function of the SINR or even the capacity per area.

\end{abstract}

\section{Introduction}

Next-generation mobile networks are expected to provide reliable connectivity to UAVs\footnote{Short for uncrewed aerial vehicles, commonly known as \emph{drones}.} for low-latency control and mission-specific data payloads \cite{GerGarAza2022,wu20205g,FotQiaDin2019}.
However, cellular base stations (BSs) are traditionally designed to optimize \emph{2D connectivity} on the ground, which results in UAVs not being reached by their primary antenna lobes. Furthermore, UAVs flying above buildings experience line-of-sight (LoS) interference from numerous BSs, causing a degraded signal-to-interference-plus-noise ratio (SINR) \cite{GerGarGal2018,ZenLyuZha2019}. 
Achieving \emph{3D connectivity} requires re-engineering existing ground-focused deployments. 
Recent proposals for ubiquitous aerial connectivity rely on network densification \cite{KanMezLoz2021,DanGarGer2020}, dedicated infrastructure for aerial services \cite{GerLopBen2022,MozLinHay2021}, or utilizing satellites to supplement the ground network \cite{BenGerLop2022}, all of which may require costly hardware or signal processing upgrades.

Pessimistic conclusions from the above stem from the assumption that UAVs will fly uncontrollably and that cellular networks must provide coverage in every 3D location. However, as the number of UAVs increases, they could be restricted to specific air routes, known as \emph{UAV corridors}, regulated by the appropriate authorities \cite{CheJaaYan2020}. 
With the concept of UAV corridors gaining acceptance, researchers have started studying UAV trajectory optimization by matching a UAV's path to the best network coverage pattern \cite{BulGuv2018,ChaSaaBet2018,EsrGanGes2020,BayTheCac2021}. However, the definition of UAV corridors will likely prioritize safety over network coverage, limiting the scope of coverage-based UAV trajectory optimization and requiring instead a 3D network design tailored to UAV corridors. 
Recent research has focused on fine-tuning cellular deployments for UAV corridors using ad-hoc system-level optimization \cite{BerLopGes2022,ChoGuvSaa2021,SinBhaOzt2021,BerLopPio2023}, as well as theoretical analysis \cite{KarGerJaf2023,KarGerJafICC2023,MaeChoGuv2021}. Despite these promising contributions, a scalable optimization framework is still needed to maximize performance functions that are mathematically intractable. 

In this paper, we propose a new methodology based on Bayesian optimization (BO) to design a cellular deployment for both ground users (GUE) and UAVs flying along corridors. For traditional ground-focused networks, BO has proven useful in achieving coverage/capacity tradeoffs \cite{Dreifuerst2020}, optimal radio resource allocation \cite{Maggi2020,zhang2023bayesian}, and mobility management \cite{CarValGer2023}. BO can effectively maximize expensive-to-evaluate stochastic performance functions, and unlike other non-probabilistic methods, converge rapidly 
without requiring a large amount of data. 
As a case study, we maximize the mean SINR perceived by GUEs as well as UAVs on corridors by optimizing the electrical antenna tilts and the transmit power employed by each BS. We do so under realistic 3GPP assumptions for the network deployment and propagation channel model. 

Our main findings can be summarized as follows:
\begin{itemize}[leftmargin=*]
    \item
    The proposed algorithm reaches convergence in less than 170 iterations for all scenarios tested. In all cases, after as few as 80\,iterations, the algorithm only falls short of its final performance by less than 10\,\%. 
    \item
    Unlike a traditional cellular configuration where all BSs are downtilted and transmit at full power, pursuing a signal quality tradeoff between the ground and the UAV corridors results in a subset of the BSs being uptilted, with the rest remaining downtilted or turned-off. Such configuration is highly non-obvious and difficult to design heuristically.
    \item 
    The proposed optimized network boosts the SINR on the UAV corridor, with a 23.4\,dB gain in mean compared to an all-downtilt, full-power baseline. Meanwhile, it nearly preserves the SINR on the ground, even attaining a gain of 1.3\,dB in mean SINR with respect to said baseline.
\end{itemize}
\section{System Model}
\label{sec:system_model}

We now introduce the deployment, channel model, and performance metric considered. 
(Also see Table~\ref{table:parameters}.)

\begin{figure}
\centering
\includegraphics[width=\figwidth]{
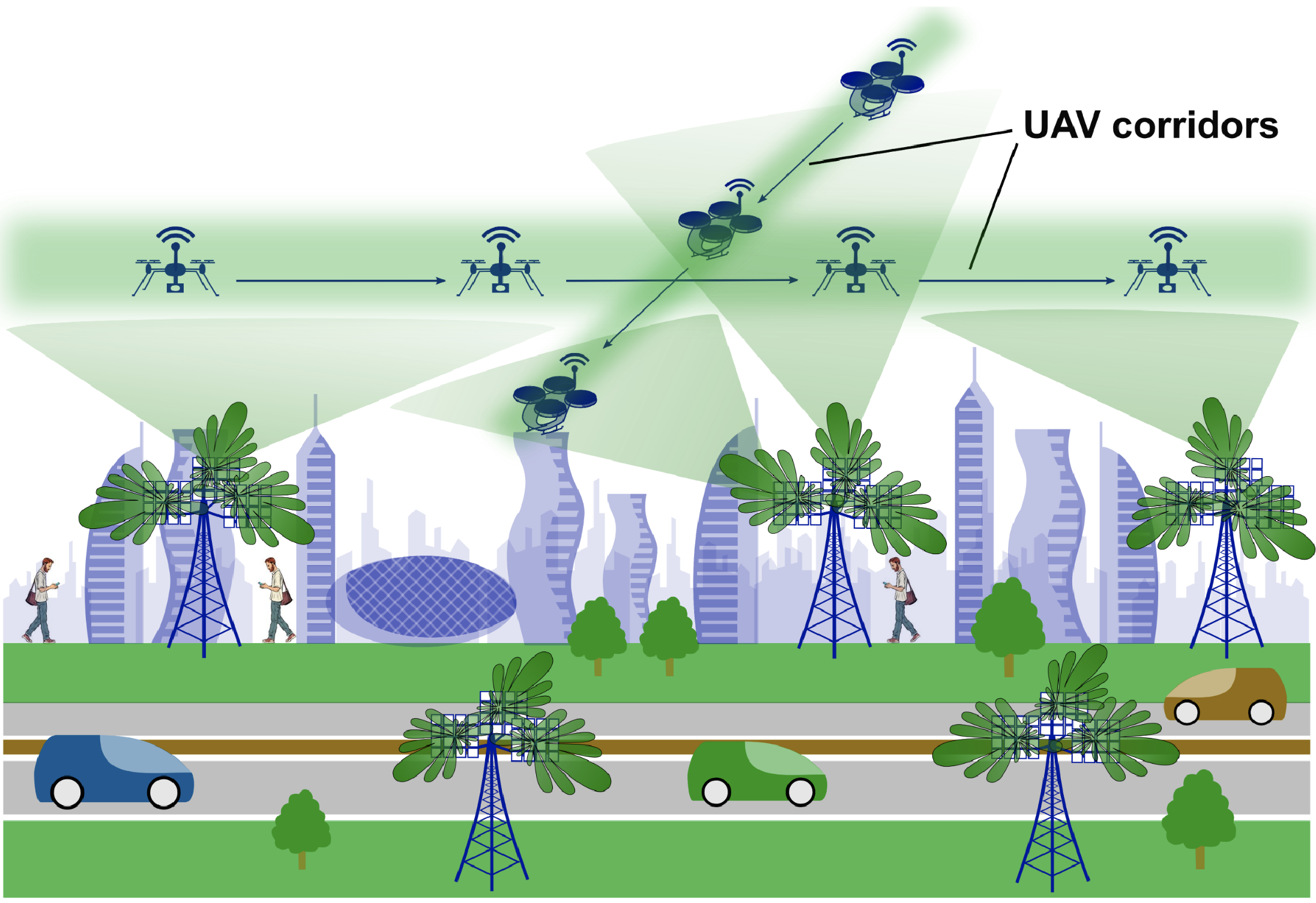}
\caption{Cellular network with downtilted and uptilted BSs supporting GUEs and UAVs flying along corridors (blurred green) \cite{KarGerJafICC2023}.}
\label{fig:illustration}
\vspace{-0.5cm}
\end{figure}

\subsection{Network Deployment}
We consider the downlink of a cellular network as specified by the 3GPP~\cite{3GPP38901,3GPP36777}. A total of 57 BSs are deployed at a height of 25\,m. BSs are deployed on a wrapped-around hexagonal layout consisting of 19 sites with a 500\,m inter-site distance (ISD). A site comprises three co-located BSs, each creating a sector (i.e., a cell) spanning a $120^{\circ}$ angle in azimuth. 
Let $\ncalB$ denote the set of BSs. We set the transmit power $p_{b} \leq 46$\,dBm and vertical antenna tilt $\theta_{b} \in [-90^{\circ},90^{\circ}]$ of each BS $b\in\ncalB$ as the object of optimization, with negative and positive angles denoting downtilts and uptilts, respectively. 
The network serves all user equipment (UE), i.e., both GUEs and UAVs, whose sets are denoted as $\ncalG$ and $\ncalU$, respectively. All GUEs are distributed uniformly across the entire cellular layout at a height of 1.5\,m, with an average of 15\,GUEs per sector. UAVs are uniformly distributed along a predefined aerial region consisting of four corridors arranged as specified in Table~\ref{table:parameters} and illustrated in Fig.~\ref{fig:UAV_Cells}, with an average of 50 uniformly deployed UAVs per corridor.

\begin{table}
\centering
\caption{System-level parameters \cite{3GPP36777, 3GPP38901}}
\label{table:parameters}
\def\arraystretch{1.2}
\begin{tabulary}{\columnwidth}{ |p{2.1cm} | p{5.85cm} | }
\hline
	\textbf{Deployment} 			&  \\ \hline
  Cellular layout				& Hexagonal grid, $\mathrm{ISD} = 500$\,m, three sectors per site, one BS per sector at $25$~m, wrap-around \\ \hline
  Frequency band 		&  10\,MHz in the 2\,GHz band \\ \hline
	BS max power 			& 46\,dBm \cite{3GPP38901}  \\ \hline   
	BS antenna 		& Vert./Horiz.  HPBW: $10^{\circ}$/$65^{\circ}$, max. gain: 8~dBi \\ \hline\hline

	\textbf{Users} 			&  \\ \hline
  GUE distribution 				& 15 per sector on average, at 1.5\,m \\ \hline
  	\multirow{5}{*}{UAV distribution}  &  Uniform in four aerial corridors with coordinates: \\ 
	 				& $[-650,-610] \times [-780,780]$ at 150\,m\\ 
	 				& $[-780,780] \times [-650,-610]$ at 120\,m \\ 
	 				& $[-780,780] \times [610,650]$ at 120\,m\\ 
	 				& $[610,650] \times [-780,780]$ at 150\,m\\
      & 50 UAVs per corridor on average (also see Fig.~\ref{fig:UAV_Cells})\\ \hline
      
	User association				& Based on RSS (large-scale fading) \\ \hline

	User antenna 		& Omnidirectional, gain: 0~dBi \\ \hline\hline

	\textbf{Channel model} 			&  \\ \hline
	Large-scale fading 		& Urban Macro as per \cite{3GPP36777, 3GPP38901} 
           \\ \hline
	Small-scale fading		& GUEs: Rayleigh. UAVs: pure LoS. \\ \hline
	Thermal noise 				& -174\,dBm/Hz density, 9\,dB noise figure \\ \hline 
\end{tabulary}
\end{table}

\subsection{Propagation Channel}

The network operates on a 10\,MHz band in the 2\,GHz spectrum, with the available bandwidth fully reused across all cells. All radio links experience path loss and lognormal shadow fading. 
BSs are equipped with a directive antenna with a maximum gain of 8\,dBi and a vertical (resp. horizontal) half-power beamwidth of 10$^{\circ}$ (resp. 65$^{\circ}$). 
All UEs are equipped with a single omnidirectional antenna. We denote $G_{b,k}$ as the large-scale power gain between BS $b$ and UE $k$, comprising path loss, shadow fading, and antenna gain, with the latter depending on the antenna tilt $\theta_{b}$. 
We denote $h_{b,k}$ as the small-scale block fading between cell $b$ and UE $k$. We assume that the GUEs undergo Rayleigh fading and that the UAV links experience pure LoS propagation conditions, given their elevated position with respect to the clutter of buildings.%
\footnote{The small-scale fading model does not affect the conclusions drawn herein.}
Each UE $k$ is associated with the BS $b_k$ providing the largest average received signal strength (RSS).

\subsection{Performance Metric}

The downlink SINR in decibels (dB) experienced by UE $k$ from its serving BS $b_k$ on a given time-frequency physical resource block is given by
\begin{equation}
  \sinr_{\textrm{dB},k} = 10\,\log_{10} \left( \,\frac{p_{b_k} \cdot G_{b_k,k} \cdot |h_{b_k,k}|^2 }{
  \sum\limits_{b\in\calB\backslash b_k}{p_{b} \cdot G_{b,k} \cdot |h_{b,k}|^2 \,+\, \sigma_{\textrm{T}}^2}}\right),
  \label{SINR_DL_TN}
\end{equation}
where $\sigma_{\textrm{T}}^2$ denotes the thermal noise power. The SINR in (\ref{SINR_DL_TN}) depends on the vertical antenna tilts $\theta_{b}$ as well as on the transmit powers $p_{b}$ of all BSs---the former through the large-scale gains $G_{b,k}, b\in\calB$.

Our goal is to determine the set of BS antenna tilts and transmit powers that maximize the downlink SINR in (\ref{SINR_DL_TN}) averaged over all UEs in the network.%
\footnote{Note that the proposed framework is amenable to maximize any desired function---not necessarily the mean---of the RSS, SINR, or the capacity.}
We therefore define the following objective function $f(\cdot)$ to be maximized:
\begin{equation}
    f\left(\bm{\theta}, \bm{p}\right) = \frac{\lambda}{\|\ncalU\|} \cdot \sum\limits_{k\in\ncalU}{\sinr_{\textrm{dB},k}} + \frac{1-\lambda}{\|\ncalG\|} \cdot \sum\limits_{k\in\ncalG}{\sinr_{\textrm{dB},k}},
    \label{eqn:Opt_problem}
\end{equation}
where the vectors $\bm{\theta}$ and $\bm{p}$ respectively contain the antenna tilts $\theta_{b}$ and transmit powers $p_b$ of all BSs $b$ $\in$ $\calB$ and $\|\cdot\|$ denotes the cardinality of a set. The parameter $\lambda\in[0,1]$ is a mixing ratio that trades off GUE and UAV performance. As special cases, $\lambda=0$ and $\lambda=1$ optimize the cellular network for GUEs only and UAVs only, respectively. 
\section{Proposed Methodology}

In this paper, we use Bayesian optimization to determine the set of BS antenna tilts and transmit powers that maximize the objective function  defined in \eqref{eqn:Opt_problem}. BO is a framework suitable for black-box optimization, where the objective function $f(\cdot)$ is non-convex, non-linear, stochastic, high-dimensional and/or computationally expensive to evaluate. 
In essence, BO uses the Bayes theorem to perform an informed search over the solution space, and works by iteratively constructing a probabilistic \textit{surrogate model} of the function being optimized based on prior evaluations of such function at a number of points in the search space \cite{shahriari2015taking}. 
The surrogate model is easier to evaluate than the function being optimized
and is updated with each point evaluated. An \textit{acquisition function} $\alpha(\cdot)$ is then used to interpret and score the response from the surrogate to decide which point in the search space should be evaluated next. 
The acquisition function balances exploration (searching for new and potentially better solutions) and exploitation (focusing on the currently best-performing solutions).
Further details on our methodology are provided as follows. 


\subsection{Evaluation of the Objective Function and Surrogate Model}

In this paper, 
a query point $\textbf{x} = [\bm{\theta}^\top, \bm{p}^\top]^\top$ is defined by a configuration of antenna tilts $\theta_{b}$ and transmit powers $p_b$ of all BSs $b$ $\in$ $\calB$. 
The corresponding value of the objective function $f(\textbf{x})$ is the mean SINR over all UEs under given antenna tilts $\bm{\theta}$ and transmit powers $\bm{p}$ and is obtained from (\ref{eqn:Opt_problem}).
For convenience, let us define $\textbf{X} = [\textbf{x}_1,\ldots,\textbf{x}_N]$ as a set of $N$ points 
and $\textbf{f}(\textbf{X})=[f_1,\ldots,f_N]^\top$ as the set of corresponding objective function evaluations, 
with $f_i = f(\textbf{x}_i)$, $i=1,\ldots,N$. 
As described in Section~\ref{sec:system_model},
the objective $f(\cdot)$ being optimized is a mathematically intractable stochastic function driven by the models and assumptions further detailed in Table~\ref{table:parameters}, which may be obtained by a cellular operator in real-time. To validate our proposed framework, we evaluate $f(\cdot)$ through system-level simulations. Such simulations are affected by the inherent randomness of the UE locations and the probabilistic channel model in (\ref{SINR_DL_TN}), 
thus yielding a noisy sample $\tilde{f}(\textbf{x})$ when evaluating a given point $\textbf{x}$.

Following a standard BO framework,
we use a Gaussian process (GP) regressor to create a surrogate model that approximates the objective function, denoted as $\widehat{f}(\cdot)$ \cite{shahriari2015taking}.
The resulting GP model allows to predict the value of $\tilde{f}(\textbf{x})$ for a queried point $\textbf{x}$ given the previous observations $\tilde{\textbf{f}}(\textbf{X})=\tilde{\textbf{f}}$ over which the model is constructed. 
Formally, 
the GP prior on the objective $\tilde{f}(\textbf{x})$ prescribes that, 
for any set of inputs $\textbf{X}$,
the corresponding objectives $\tilde{\textbf{f}}$ are jointly distributed as
\begin{equation}
  p(\,\tilde{\textbf{f}}\,) = \mathcal{N}(\,\tilde{\textbf{f}} \,|\, \boldsymbol{\mu}(\textbf{X}),\mathbf{K}(\textbf{X})\,),
  \label{posterior}
\end{equation}
where $\boldsymbol{\mu}(\textbf{X}) = [\mu(\mathrm{\textbf{x}}_1),\ldots,\mu(\mathrm{\textbf{x}}_N)]^\top$ is the $N \times 1$ mean vector, 
and $\mathbf{K}(\textbf{X})$ is the $N \times N$ covariance matrix, 
whose entry $(i,j)$ is given as $[\textbf{K}(\textbf{X})]_{i,j} = k(\textbf{x}_{i},\textbf{x}_{j})$ with $i,j \in \{1,\hdots,N\}$. 
For any point $\textbf{x}$, 
the mean $\mu(\textbf{x})$ provides a prior knowledge on the objective $f(\textbf{x})$, 
while the kernel $\mathbf{K}(\textbf{X})$ indicates the uncertainty across pairs of values of \textbf{x}. 
%

Given a set of observed noisy samples $\tilde{\textbf{f}}$ at previously sampled points $\textbf{X}$, 
the posterior distribution of $\widehat{f}(\textbf{x})$ at point $\textbf{x}$ can be obtained as
\begin{equation}
  p(\widehat{f}(\textbf{x}) = \widehat{f} \, | \, \textbf{X},\textbf{$\tilde{\textbf{f}}$} \,) = \mathcal{N}(\widehat{f} \,|\, \mu(\textbf{x} \,|\, \textbf{X},\tilde{\textbf{f}}),\sigma^2(\textbf{x} \,|\, \textbf{X},\tilde{\textbf{f}})),
  \label{posterior_Noisy}
\end{equation}
with
\begin{equation}
  \mu(\textbf{x} \,|\, \textbf{X},\tilde{\textbf{f}}) = \mu(\textbf{x}) + \tilde{\textbf{k}}(\textbf{x})^\top (\tilde{\textbf{K}}(\textbf{X}))^{-1}(\tilde{\textbf{f}}-\boldsymbol{\mu}(\textbf{X})),
  \label{Mean_posterior_Noisy}
\end{equation}
\begin{equation}
  \sigma^2(\textbf{x} \,|\, \textbf{X},\tilde{\textbf{f}}) = k(\textbf{x},\textbf{x}) - \tilde{\textbf{k}}(\textbf{x})^\top (\tilde{\textbf{K}}(\textbf{X}))^{-1} \,\tilde{\textbf{k}}(\textbf{x}),
  \label{Kernel_posterior_Noisy}
\end{equation}
where 
$\tilde{\textbf{k}}(\textbf{x}) = [k(\textbf{x},\textbf{x}_{1}),\ldots,k(\textbf{x},\textbf{x}_{N})]^\top$ is the $N \times 1$ covariance vector and $\tilde{\textbf{K}}(\textbf{X}) = \textbf{K}(\textbf{X}) + \sigma^2 \textbf{I}_{\text{N}}$, with $\sigma^2$ denoting the observation noise represented by the variance of the Gaussian distribution, and $\textbf{I}_{\text{N}}$ denoting the $N \times N$ identity matrix.
Note that \eqref{Mean_posterior_Noisy} and \eqref{Kernel_posterior_Noisy} represent the mean and variance of the estimation, the latter indicating the degree of confidence.  

\subsection{Proposed BO Algorithm and Acquisition Function}
The proposed BO algorithm starts by creating a GP prior $\{\mu(\cdot), k(\cdot, \cdot)\}$ based on a dataset $\mathcal{D} = \{\textbf{x}_1,\ldots,\textbf{x}_{N_{\textrm{o}}},\tilde{f}_1,\ldots,\tilde{f}_{N_{\textrm{o}}}\}$ containing $N_{\textrm{o}}$ initial observations. The dataset is constructed via system-level simulations according to the model and objective function defined in Section~\ref{sec:system_model}. The antenna tilts $\bm{\theta}_i$ and transmit powers $\bm{p}_i$ for every observation point $\textbf{x}_i = [\bm{\theta}_i^\top, \bm{p}_i^\top]^\top$ in $\mathcal{D}$ are chosen randomly in $[-90^{\circ}, 90^{\circ}]$ and $[6\,\text{dBm}, 46\,\text{dBm}]$, respectively.

Once the initial GP prior is constructed, the vectors $\bm{\theta}_0$ and $\bm{p}_0$ are initialized with all entries set to $0^{\circ}$ and $46\,\text{dBm}$, respectively. We denote $\tilde{f}^{*}$ as the best observed objective value, which is initialized to $\tilde{f}_0^{*} = -\infty$. 
The algorithm then iterates over each BS $b \in \calB$, one at a time.%
\footnote{At iteration $n$, the BS considered is thus $b_{n} = ((n-1) \mod \|\ncalB\|) + 1$.}
At each such iteration $n$, 
only the antenna tilt and transmit power of the BS $b_{n}$ under consideration
are updated, while keeping the remaining entries of $\bm{\theta}_n$ and $\bm{p}_n$ fixed to their values from the previous iteration. The query point under optimization is thus reduced to a two-dimensional vector that we will denote as $\widehat{\textbf{x}}_{n} = [\theta_{b_{n}}, p_{b_{n}}]$. 

The algorithm then leverages the observations in $\mathcal{D}$ to choose $\widehat{\textbf{x}}_{n}$. This is performed via an acquisition function $\alpha(\cdot)$, which is designed to trade off the exploration of new points in less favorable regions of the search space with the exploitation of well-performing ones. 
The former prevents getting caught in local maxima, 
whereas the latter minimizes the risk of excessively degrading performance.%
\footnote{While in this paper we run the proposed optimization on system-level simulations, 
its practical implementation requires testing the performance (mean SINR) of each candidate point (BS tilt and power) in a real network, 
whereby it becomes undesirable to explore poorly performing points.}
In what follows, we adopt the expected improvement (EI) as the acquisition function, which has shown to perform well in terms of balancing the trade-off between exploration and exploitation \cite{shahriari2015taking, Maggi2020}. At every iteration $n$, the EI tests and scores a set of $N_{\textrm{c}}$ randomly drawn candidate points $\{\widehat{\textbf{x}}_{\text{cand}_{1}},\ldots,\widehat{\textbf{x}}_{\text{cand}_{N_{\textrm{c}}}}\}$ through the surrogate model. The EI is defined as \cite{huang2006sequential,Maggi2020} 
\begin{equation}
    \begin{aligned}
    \alpha \left(\widehat{\textbf{x}}_{\text{cand}} \,|\, \mathcal{D} \right) =\, & [\, \mu\,(\widehat{\textbf{x}}_{\text{cand}} \,|\, \mathcal{D})\, - \widehat{f}^{*} - \xi] \cdot \Phi(\delta)\\
    & +\sigma^2\,(\widehat{\textbf{x}}_{\text{cand}} \,|\, \mathcal{D}) \cdot \phi(\delta),
    \label{eqn:EI}
    \end{aligned}
\end{equation}
where $\widehat{f}^{*} = 
{\textrm{max}_i}\,\{\widehat{f}_{\text{cand}_{i}}\}$ denotes the current best approximated objective value according to the surrogate model, $\Phi$ (resp. $\phi$) is the standard Gaussian cumulative (resp. density) distribution function, and
\begin{equation}
    \delta = \frac{\mu\,(\widehat{\textbf{x}}_{\text{cand}} \,|\, \mathcal{D})\, - \widehat{f}^{*} - \xi}{\sigma^2\,(\widehat{\textbf{x}}_{\text{cand}} \,|\, \mathcal{D})\,},
    \label{eqn:EI_delta}
\end{equation}
with $\mu(\widehat{\textbf{x}}_{\text{cand}}\,|\, \mathcal{D})$ and $\sigma^2\,(\widehat{\textbf{x}}_{\text{cand}} \,|\, \mathcal{D})\,$ given in \eqref{Mean_posterior_Noisy} and \eqref{Kernel_posterior_Noisy}, respectively. 
The parameter $\xi\in[0,1)$ in (\ref{eqn:EI}) and (\ref{eqn:EI_delta}) regulates the exploration vs. exploitation tradeoff, with larger values promoting the former, and vice versa. In this paper, we aim for a risk-sensitive EI acquisition function and set $\xi = 0.01$.

Leveraging batch evaluation, which allows for automatic dispatch of independent operations across multiple computational resources (e.g., GPUs), at each iteration we evaluate a set of $N_{\textrm{c}}=500$ candidate points through the surrogate model, using 10 batches each consisting of 50 points. The query
point $\widehat{\textbf{x}}_{n}$ is then chosen as
\begin{equation}
    \widehat{\textbf{x}}_{n} = \underset{\substack{i}}{\textrm{arg}\,\textrm{max}}\;\; \alpha \left(\widehat{\textbf{x}}_{\text{cand}_i} \,|\, \mathcal{D} \right).
    \label{eqn:BO_acq}
\end{equation}
Once $\widehat{\textbf{x}}_{n} = [\theta_{b_{n}}, p_{b_{n}}]$ is determined, the vectors $\bm{\theta}_n$ and $\bm{p}_n$ are obtained from $\bm{\theta}_{n-1}$ and $\bm{p}_{n-1}$ by replacing their $b_{n}$-th entries with $\theta_{b_{n}}$ and $p_{b_{n}}$, respectively, yielding $\textbf{x}_n = [\bm{\theta}_n^\top, \bm{p}_n^\top]^\top$. A new observation of the objective function $\tilde{f}(\textbf{x}_n)$ is then produced, and the dataset $\mathcal{D}$, the GP prior, and the best observed objective value $\tilde{f}^{*}$ are all updated accordingly. The algorithm then moves on to optimizing the antenna tilt and transmit power of BS $b_{n+1}$, until all BSs in $\calB$ have been optimized. 
This loop over all BSs is then repeated until the best observed value $\tilde{f}^{*}$ has remained unchanged for $\ell_{\text{max}}$ consecutive loops, after which the algorithm recommends the point $\textbf{x}^{*}$ that produced the best observation $\tilde{f}^{*}$. The proposed approach is summarized in Algorithm~\ref{alg:BO}.

\begin{algorithm}[hbt!]
\caption{Proposed BO algorithm}\label{alg:BO}
\textbf{Input:} Initial dataset $\mathcal{D} = \{\textbf{x}_1,\ldots,\textbf{x}_{N_{\textrm{o}}},\tilde{f}_1,\ldots,\tilde{f}_{N_{\textrm{o}}}\}$;

\textbf{Output:} Optimal configuration $\textbf{x}^{*}$;

\textbf{Initialization:}

\quad Create a GP prior $\{\mu(\cdot), k(\cdot, \cdot)\}$ using $\mathcal{D}$ and (\ref{posterior});

\quad Set all entries of $\bm{\theta}_0$ to $0^{\circ}$ and all entries of $\bm{p}_0$ to $46~\text{dBm}$;


\quad Set $\textbf{x}_0 = [\bm{\theta}_0^\top, \bm{p}_0^\top]^\top$, $n = \ell=1$, $\ell_{\textrm{max}} = 3$, $\tilde{f}^{*} = \tilde{f}_0^{*} = -\infty$;


\While{$\ell \leq \ell_{\emph{max}}$}{
$b_{n} = ((n-1) \mod \|\ncalB\|) + 1$; 

Draw $N_{\textrm{c}}$ random candidate points $\{\widehat{\textbf{x}}_{\text{cand}_{1}},\ldots,\widehat{\textbf{x}}_{\text{cand}_{N_{\textrm{c}}}}\}$;

Evaluate all candidate points using \eqref{eqn:EI};

Obtain $\widehat{\textbf{x}}_{n} = [\theta_{b_{n}}, p_{b_{n}}]$ from \eqref{eqn:BO_acq};

Update $\textbf{x}_{n}$ with $\theta_{b_{n}}$ and $p_{b_{n}}$;

Obtain observation $\tilde{f}_n = \tilde{f}(\textbf{x}_n)$ using \eqref{eqn:Opt_problem};

Augment $\mathcal{D}$ by including $\textbf{x}_{n}$ and $\tilde{f}_n$;

Update the GP prior $\{\mu(\cdot), k(\cdot, \cdot)\}$ using $\mathcal{D}$ and (\ref{posterior});

\If{$\tilde{f}_{n} > \tilde{f}^{*}_{n-1}$}
{$\tilde{f}_n^{*} \gets\tilde{f}_{n}$;
$\textbf{x}_{n}^{*} \gets \textbf{x}_{n}$;
}
\Else{$\tilde{f}_n^{*} \gets \tilde{f}^{*}_{n-1}$;
}

\If{$b_n=\|\ncalB\|$}
{
\If{$\tilde{f}^{*}_{n} > \tilde{f}^{*}$}
{$\tilde{f}^{*} \gets \tilde{f}^{*}_{n}$;
$\textbf{x}^{*} \gets \textbf{x}_{n}^{*}$;
$\ell \gets 0$;
}
{$\ell \gets \ell + 1$;
}}
$n \gets n + 1$;
}
\end{algorithm}

\section{Numerical Results}

\begin{figure}
\centering
\includegraphics[width=\figwidth]{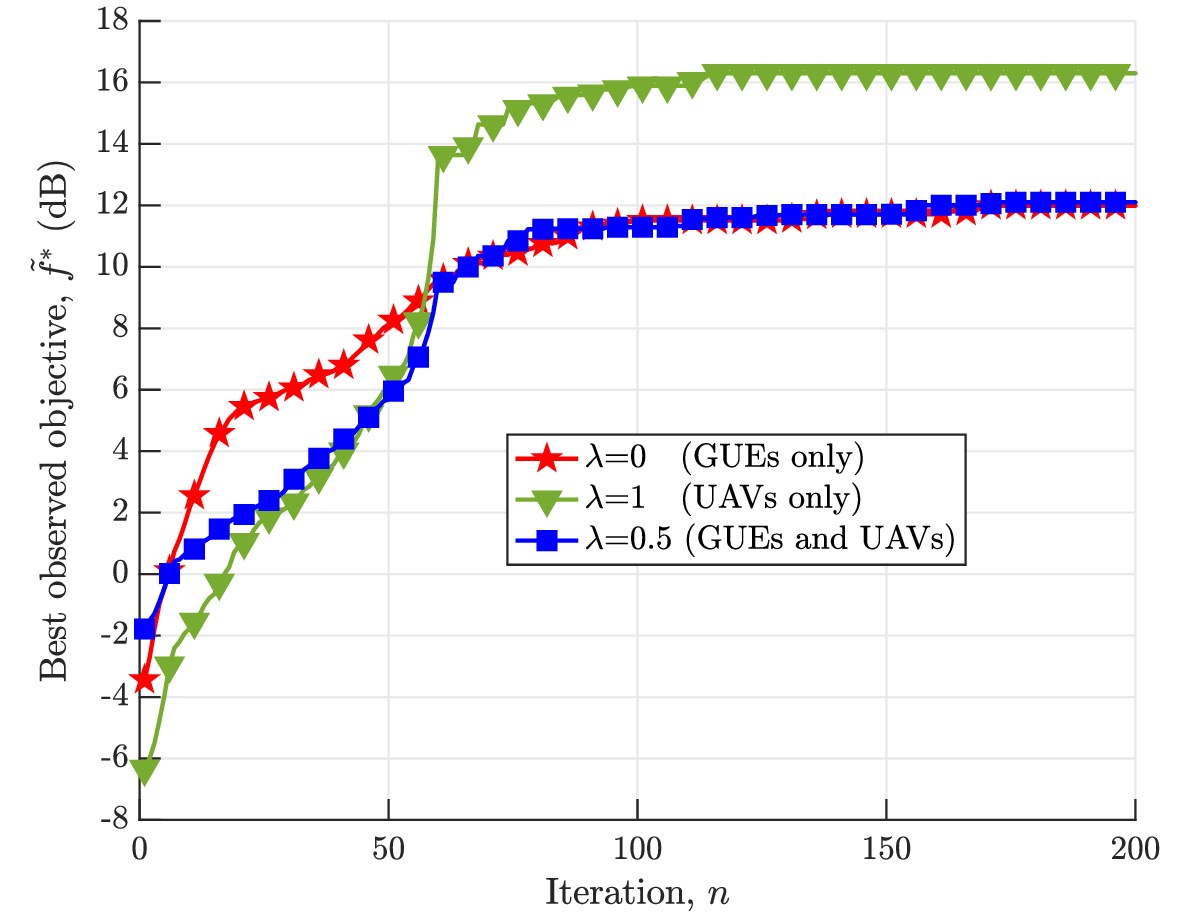}
\caption{Convergence of the proposed algorithm, showing the evolution of the best observed objective vs. the number of iterations $n$.}
\label{fig:BO_Convergence}
\end{figure}

In this section, 
we present the results obtained when applying our proposed framework introduced in Section III on the system model defined in Section II, 
for three values of $\lambda$, namely 0, 1, and 0.5. 
We recall that these values correspond to optimizing the cellular network for GUEs only, for UAVs only, and for both with equal weight, respectively. 
The BO algorithm is run on BoTorch, an open-source library built upon PyTorch \cite{balandat2020botorch}. 
We use the Matern-5/2 kernel for $\mathbf{K}(\textbf{X})$ and fit the GP hyperparameters using maximum posterior estimation. 

\subsubsection*{Convergence of the BO framework} 

Fig.~\ref{fig:BO_Convergence} shows the convergence of the proposed BO algorithm by illustrating the best observed objective at each iteration $n$. 
Convergence is reached in less than 170 iterations for all three values of $\lambda$. 
In all cases, after as few as 80 iterations the algorithm only falls short of its final performance by less than 10\%. In the remainder of this section, we discuss the network configuration recommended by the algorithm and quantify its final performance.

\subsubsection*{Optimal network configuration} 

\begin{figure}
\centering
\includegraphics[width=\figwidth]{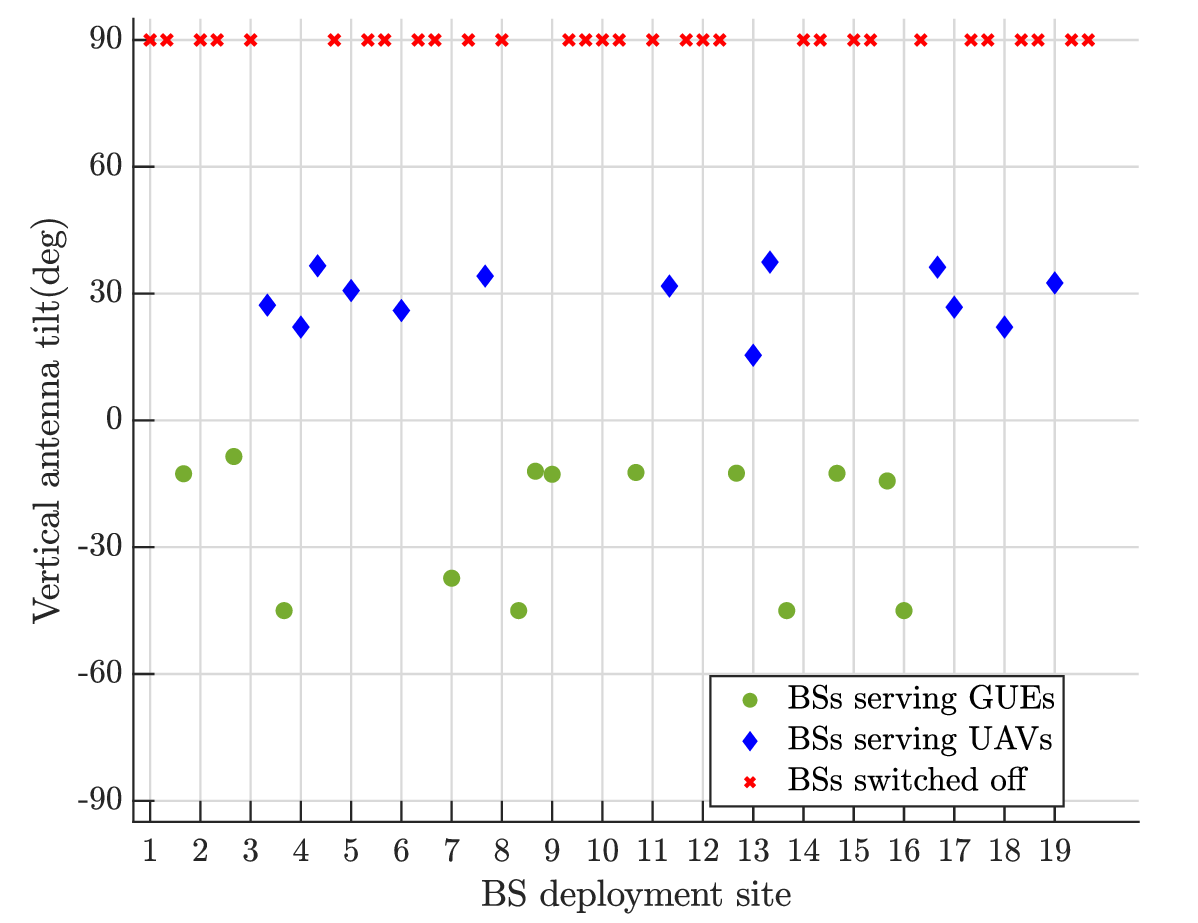}
\caption{Optimized BS tilts for both GUEs and UAVs ($\lambda = 0.5$). Green circles, blue diamonds, and red crosses respectively denote BSs serving GUEs, serving UAVs, and switched off.}
\label{fig:Opt_values_tilt}
\end{figure}

\begin{figure}
\centering
\includegraphics[width=\figwidth]{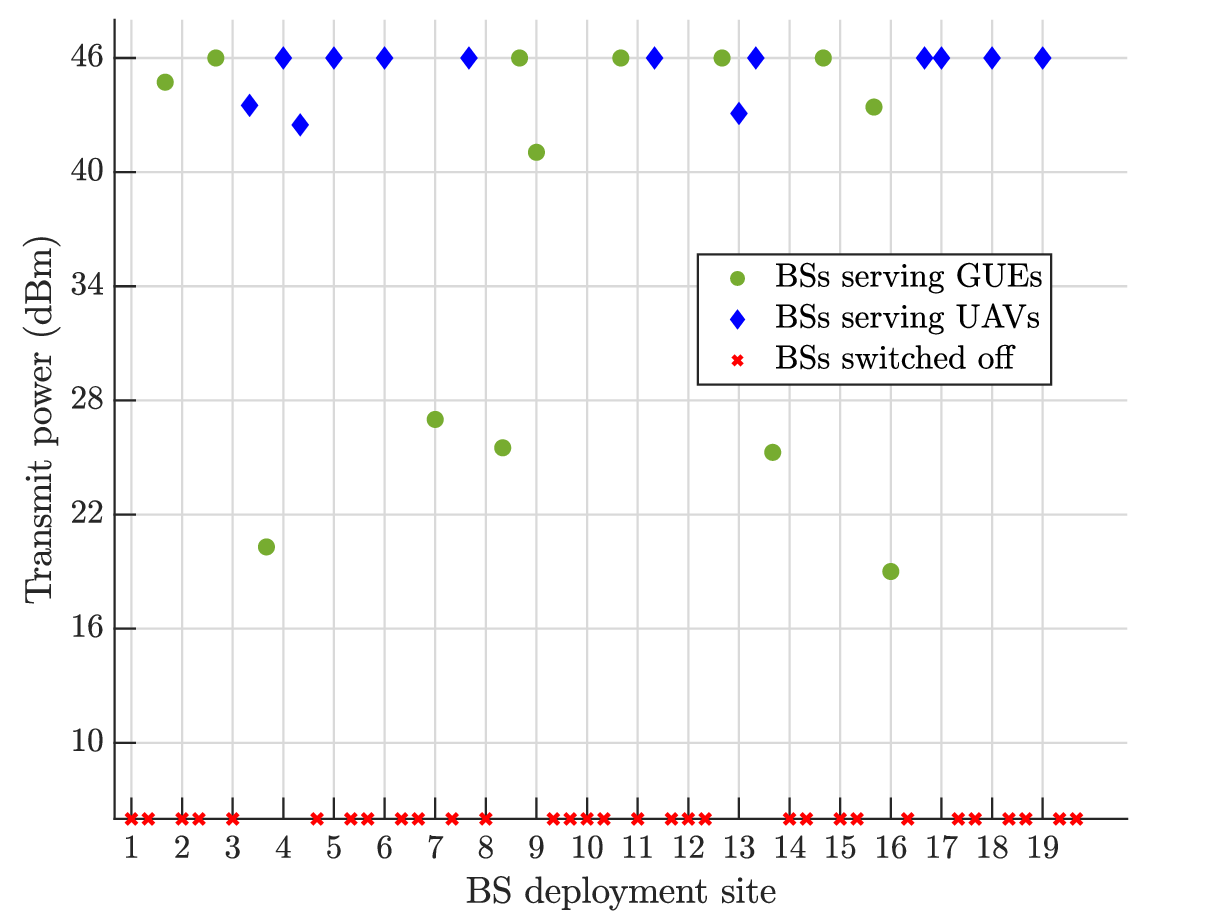}
\caption{Optimized BS power for both GUEs and UAVs ($\lambda = 0.5$).}
\label{fig:Opt_values_power}
\end{figure}

Fig.~\ref{fig:Opt_values_tilt} and Fig.~\ref{fig:Opt_values_power} respectively show the optimal values of the vertical electrical antenna tilts $\bm{\theta}$ and transmit powers $\bm{p}$ for the case $\lambda=0.5$, where a tradeoff is sought between SINR on the ground and along the aerial corridors.
In both figures, the BS index denotes the deployment site (black dots in Fig.~\ref{fig:UAV_Cells}), each comprising three sectors (cells). Markers indicate whether each cell is serving GUEs (green circles), UAVs (blue diamonds), or it is switched off to mitigate unnecessary interference (red crosses). 
The figures show that, unlike a traditional cellular network configuration where all BSs are downtilted (e.g., to $-12^{\circ}$ \cite{3GPP38901}) and transmit at full power, 
pursuing an SINR tradeoff between the ground and the UAV corridors results in a subset of the BSs being uptilted (i.e., a total of 13 BSs), 
with the rest remaining downtilted or turned off. 
Such configuration is non-obvious and would be difficult to design heuristically.

\subsubsection*{Connectivity along UAV corridors} 

\begin{figure}
\centering
\includegraphics[width=\figwidth]{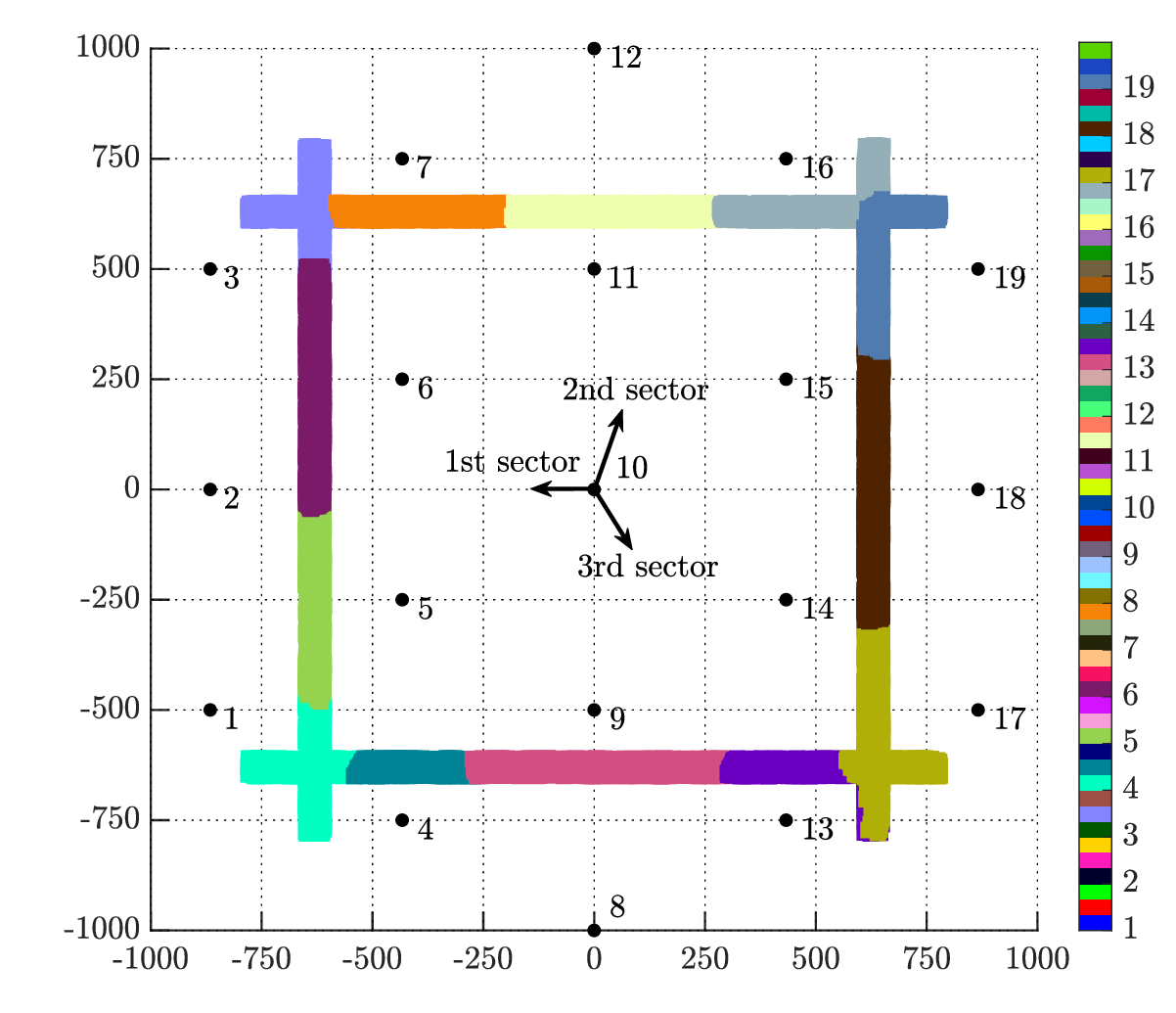}
\caption{Cell partitioning for UAV corridors when the cellular network is optimized for both GUEs and UAVs ($\lambda$ = 0.5).}
\label{fig:UAV_Cells}
\end{figure}

Fig.~\ref{fig:UAV_Cells} shows the resulting cell partitioning for the UAV corridors when the network is optimized for both populations of UEs with the recommended values for BS tilts and transmit powers given in Fig.~\ref{fig:Opt_values_tilt} and Fig.~\ref{fig:Opt_values_power} for $\lambda = 0.5$.
Note that only the 13 up-tilted BSs (blue diamonds in Fig.~\ref{fig:Opt_values_tilt}) are exploited to provide service along the UAV corridors, each covering a different segment according to their geographical location and orientation. 
%

\subsubsection*{Resulting SINR performance} 

\begin{figure}
\centering
\includegraphics[width=\figwidth]{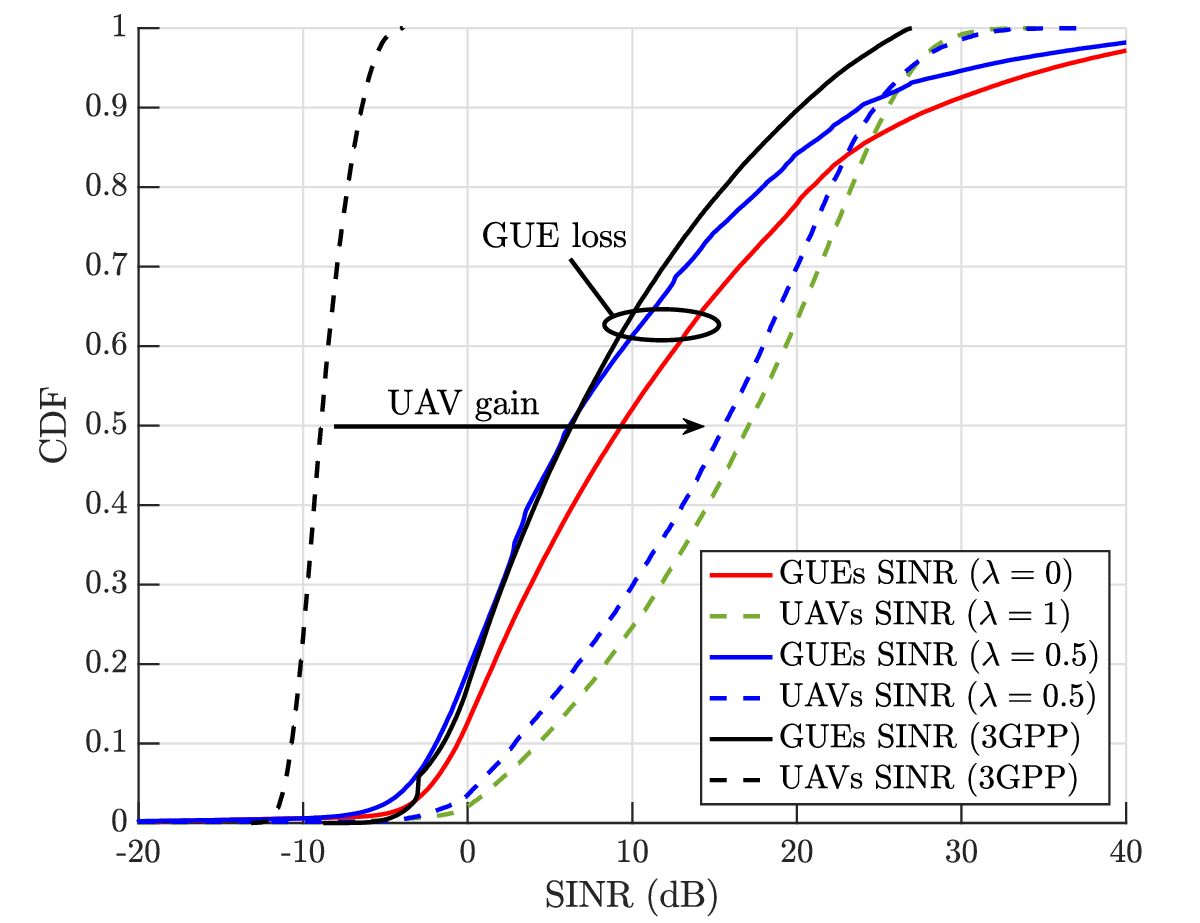}
\caption{SINR for UAVs (dashed) and GUEs
(solid) when the network is optimized for GUEs only ($\lambda$ = 0), UAVs only ($\lambda$ = 1), and both ($\lambda$ = 0.5), and for an all-downtilt, full-power baseline (3GPP).}
\label{fig:SINR_CDFs}
\end{figure}

Fig.~\ref{fig:SINR_CDFs} shows the cumulative distribution function (CDF) of the SINR perceived by GUEs (solid lines) and UAVs (dashed lines) when the cellular network is optimized for GUEs only ($\lambda = 0$, red), UAVs only ($\lambda = 1$, green), and both ($\lambda = 0.5$, blue). 
The performance of a traditional cellular network (black) is also shown as a baseline for comparison, 
where all BSs are downtilted to $-12^{\circ}$ and transmit at full power as per 3GPP recommendations \cite{3GPP38901}.  
In the sequel, we provide further tips to easily interpret Fig.~\ref{fig:SINR_CDFs}:
\begin{itemize}[leftmargin=*]
\item 
The curves labeled as \{GUE, $\lambda = 0$\} (solid red) and \{UAV, $\lambda = 1$\} (dashed green) can be regarded as performance upper bounds for GUEs and for UAVs. This is performance achieved when BS tilts and powers are optimized for mean SINR at GUEs only and UAVs only, respectively.
\item
The curves for $\lambda = 0.5$ (solid and dashed blue) show the optimal tradeoff reached by the proposed BO framework when the cellular network is designed to cater for both GUEs and UAV corridors, with equal weight.
\end{itemize}

\noindent Fig.~\ref{fig:SINR_CDFs} demonstrates that the proposed framework can optimize the cellular network in a way that significantly boosts the UAV SINR, with a 23.4 dB gain in mean compared to the all-downtilt, full-power baseline (dashed blue vs. dashed black). 
The UAV SINR even approaches the upper bound obtained when the network disregards the performance on the ground, falling short by only 1.2~dB in mean (dashed blue vs. dashed green).
At the same time, the solution nearly preserves the GUE SINR (solid blue), incurring a loss of 2.6~dB in mean with respect to the upper bound (solid red). 

When compared to the 3GPP all-downtilt, full-power baseline \cite{3GPP38901} (solid black), the optimal solution even attains a gain of 1.3~dB in mean GUE SINR. 
Indeed, said baseline was not designed for SINR, but rather for spatial reuse and capacity. It should thus come as no surprise that it slightly underperforms the proposed framework in terms of mean SINR.

\section{Conclusion}

In this paper, we proposed a new methodology to design a cellular deployment for both ground and aerial service based on Bayesian optimization.

\subsubsection*{Summary of results}
As a case study, we maximized the mean SINR perceived by GUEs as well as UAVs on corridors by optimizing the electrical antenna tilts and the transmit power employed at each BS. Unlike a traditional cellular network configuration in which all BSs are downtilted and transmit at full power, pursuing a signal quality tradeoff between the GUEs and UAVs on corridors results in a subset of the BSs being uptilted, with the rest remaining downtilted or turned off. Under this setting, our algorithm finds an optimal configuration that significantly boosts the UAV SINR, with a 23.4 dB gain in mean compared to an all-downtilt, full-power baseline. Meanwhile, this tradeoff nearly preserves the performance on the ground, even attaining a gain of 1.3~dB in mean SINR with respect to said baseline.

\subsubsection*{Future research directions}

Thanks to its ability to optimize intractable stochastic functions, the proposed framework is amenable to maximize other objectives of interest, such as an arbitary function of the RSS, SINR, or the channel capacity. In particular, we conjecture that maximizing the capacity per area would lead to a different network configuration than the one obtained for the present case study. 
Furthermore, while in this article we defined a single objective function capturing the performance on the ground and along UAV corridors, an extension of this work could consider multi-objective BO by defining separate performance functions for the GUEs and UAVs on corridors. The goal would then be the one of finding the Pareto front: a set of non-dominated solutions such that no objective can be improved without deteriorating another.

\bibliographystyle{IEEEtran}
\bibliography{journalAbbreviations, main}

\end{document}